\documentclass[prb,twocolumn,showpacs,preprintnumbers,amsmath,amssymb]{revtex4}

\usepackage{graphicx}
\usepackage{dcolumn}
\usepackage{bm}

\begin{document}

\preprint{APS/123-QED}

\title{Fano effect and Kondo effect in quantum dots formed in strongly coupled quantum wells}

\author{A. W. Rushforth}
 \altaffiliation[Present address ]{School of Physics and Astronomy, University of Nottingham, Nottingham NG7 2RD.}
 \email {Andrew.Rushforth@nottingham.ac.uk}
\author{C. G. Smith}
\author{I. Farrer}
\author{D. A. Ritchie}
\author{G. A. C. Jones}
\author{D. Anderson}
\author{M. Pepper}
 \affiliation{Cavendish Laboratory, Madingley Road, Cambridge, CB3 0HE, England
}

\date{\today}

\begin{abstract}
We present lateral transport measurements on strongly, vertically coupled quantum dots formed in separate quantum wells in a  GaAs/Al$_{x}$Ga$_{1-x}$As heterostructure. Coulomb oscillations are observed forming a honeycomb lattice consistent with two strongly coupled dots. When the tunnel barriers in the upper well are reduced we observe the Fano effect due to the interfering paths through a resonant state in the lower well and a continuum state in the upper well. In both regimes an in plane magnetic field reduces the coupling between the wells when the magnetic length is comparable to the centre to centre separation of the wells. We also observe the Kondo effect which allows the spin states of the double dot system to be probed.   \end{abstract}

\pacs{73.63.Kv, 73.23.Hk, 73.21.La, 72.15.Qm}
\maketitle
The transport properties of semiconductor quantum dots have revealed many physical phenomena including atomic-like energy spectra\cite{Tarucha}, Coulomb blockade\cite{McEuen}, the Kondo effect\cite{Goldhaber,Goldhaber2} and the Fano effect\cite{Gores,Kobayashi}. Such structures have been fabricated in GaAs/Al$_{x}$Ga$_{1-x}$As heterostructures using surface gate electrodes or by etching columnar mesas in double barrier structures containing InGaAs wells. These designs have been extended to form coupled double dot systems in which the transport properties have revealed evidence of capacitive coupling and tunnel coupling between the dots which can be controlled by the application of a gate voltage\cite{Livermore} or magnetic field\cite{Ancilotto}. More recently, the observation of the Kondo effect in double dot systems has been interpreted as evidence of spin entanglement between the excess electrons on each dot\cite{Chen,Jeong}. Coupled quantum dot systems have been identified as promising candidates to act as qubits for quantum computation\cite{Burkard}.

Here we present a study of the transport properties of a strongly coupled double dot system in which many of the above mentioned phenomena are observed. Each quantum dot is formed in a separate 150\AA GaAs quantum well in a GaAs/Al$_{x}$Ga$_{1-x}$As heterostructure. The quantum dots are defined using surface gate electrodes (Fig. 1a) and are separated by a 35\AA Al$_{0.33}$Ga$_{0.67}$As barrier. The structure includes Si doping above and below the quantum wells offset by 600\AA and 800\AA Al$_{0.33}$Ga$_{0.67}$As spacer layers respectively, resulting in a total electron density of n$_{s}$ = $2.5\times10^{11}$ cm$^{-2}$, with an average mobility,  $\mu$ = $1.2\times10^{6}$ cm$^{2}$V$^{-1}$s$^{-1}$. Ohmic contacts are formed by annealing Au, Ni and Ge into the structure. This results in a simultaneous contact to both quantum wells and electron transport can be measured laterally through both dots in parallel. This is a significant advance on previous designs incorporating epitaxial barriers because it allows the entrance and exit barriers to the dots to be controlled by a gate voltage. 
\begin{figure}[t]
\includegraphics{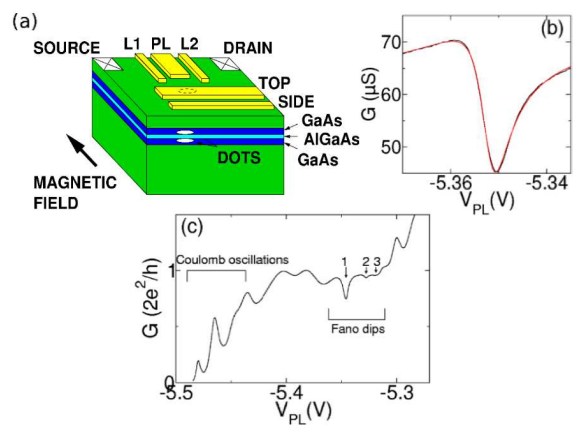}
\caption{\label{fig1:epsart} (Color online) (a) Schematic diagram of the device. The lithographic width of the dot is 500nm. (b) An example of a dip in the conductance (black) with a fit of Eq. 1 (red (dark gray)). (c) The conductance measured as a function of V$_{PL}$ for V$_{TOP}$=-0.05V (Line labelled I in Fig. 3(a)). Here V$_{L1}$, V$_{L2}$, V$_{PL}$, and V$_{SIDE}$ are swept together with V$_{L1}$ and V$_{L2}$ offset by -0.1V and V$_{SIDE}$ offset by +4V.}
\end{figure}

Transport measurements were carried out using standard ac lockin techniques with a 5$\mu$V excitation at 330Hz. All measurements were made in a dilution refrigerator at a base temperature of 50mK\cite{electrontemp}, unless stated otherwise. First, we characterised the device by forming a 1D channel by sweeping gates L1,L2,PL and SIDE simultaneously. By offsetting V$_{SIDE}$ with respect to the other three gates it was possible to shift the centre of the 1D channel such that it was positioned beneath gate TOP. This was verified by studying the conductance of the channel as a function of V$_{TOP}$ and was done so that V$_{TOP}$ could be used to vary the electron densities between the wells and to avoid the formation of conducting paths along the side of gate TOP. Following the analysis in refs. \cite{Thomas, Castleton} we can deduce that the densities of the two quantum wells are matched for V$_{TOP}$$\approx$-0.17V. Next, V$_{L1}$ and V$_{L2}$ were made slightly more negative to form tunnel barriers at the entrance and exit of a quantum dot. Figure 1(c) shows a typical conductance measurement obtained by sweeping gates L1,L2,PL and SIDE together in this regime. The features near pinch-off are regular Coulomb oscillations where peaks in the conductance represent a change of the average electron number on the dot by one. These features also form the familiar Coulomb diamond pattern when dI/dV$_{sd}$ is plotted with a dc bias, V$_{sd}$ applied between source and drain (Fig. 2(a)). Gates V$_{L1}$ and V$_{L2}$ were tuned such that the sharpest Coulomb oscillations were observed. At less negative gate voltages new features appear when the conductance reaches 2e$^{2}$/h. These are dips in the conductance which are asymmetric in gate voltage. The lineshape of these features is fitted well by the Fano formula\cite{Fano} (Fig. 1(b))

\begin{equation}
G=G_{0}+G_{1}V_{PL}+A\frac{(\epsilon+q)^2}{\epsilon^2+1}, \epsilon=\frac{\alpha(V_{PL}-V_0)}{\Gamma/2}
\end{equation}

The Fano effect arises due to interference between two paths that the electron can traverse. One is through a continuum and the other is through a resonant state at position V$_{0}$ with width $\Gamma$. The parameter q is determined by the ratio of the transmission amplitudes from an initial state i, through the discrete state $\Phi$, and continuum state $\psi$$_{E}$; q$=$$<$$\Phi$$\mid$T$\mid$i$>$/$<$$\psi$$_{E}$$\mid$T$\mid$i$>$$\pi$V$_{E}$ (V$_{E}$ is the coupling strength between these states). For q$\to$0 transmission through the continuum is dominant and the lineshape becomes a symmetric dip. The sign of q reflects the phase shift between the two paths. In Eq.1 the terms G$_{0}$ and G$_{1}$ reflect a non interfering contribution to the conductance. This arises from transmission through a 1D subband in the upper well. The dips are due to interference between this subband and resonant states in the lower well that are formed by tunnel barriers formed in that well by gates L1 and L2. It is possible for such a situation to arise when V$_{TOP}$$>$-0.17V because the electron density in the upper well will be larger than that in the lower well. This situation is depicted in Fig. 2(c). The dips form a diamond pattern (Fig. 2(b)) similar to that formed by Coulomb oscillations, supporting the conclusion that the transmission is partly through a resonant state. The increasing size of these diamonds as the gate voltage is made more negative reflects the decrease in the capacitance between the dot in the lower well and the source and drain reservoirs in that well as the tunnel barriers become more opaque. The origin of the faint lines parallel to the V$_{PL}$ axis is not clear. They may be due to excited states on the dot or weak resonances in the 1D channel in the upper well. At more negative voltages in Fig. 1(c) the transmission through the lower well is blocked and a dot is formed in the upper well giving rise to Coulomb oscillations in the conductance through that well\cite{Isolated}. The fact that the diamonds in figures 2(a) and (b) have different charging energies supports the conclusion that they arise from the conductance through different dots. This picture is consistent with the dependence of the features on V$_{TOP}$ shown in Fig. 3(a). The Fano dips have a different gradient to the Coulomb oscillations reflecting a different capacitance with gate TOP\cite{Explain shift}. Therefore, we associate these features with the lower and upper wells respectively. These features anticross in the region V$_{TOP}$=-0.15V to -0.2V which is the voltage range where the densities of the wells are known to cross. Also, in this region, the Fano dips move to lower conductance values and begin to resemble Coulomb oscillations. In this region of gate voltage it appears that we are measuring the conductance through two dots, one formed in each well. This situation is depicted in Fig. 2(d). With the application of an in plane magnetic field of 5T the Coulomb oscillations form a honeycomb pattern which has been associated with conductance through double dots \cite{Waugh}.The regions marked A-E in fig. 3(b) represent "triple points". The separation of these points is determined by the capacitive coupling between two dots and also the tunnel coupling, t giving rise to symmetric and antisymmetric states, for which the coupling strength is dependent on the in plane magnetic field.

\begin{figure}[t]
\includegraphics{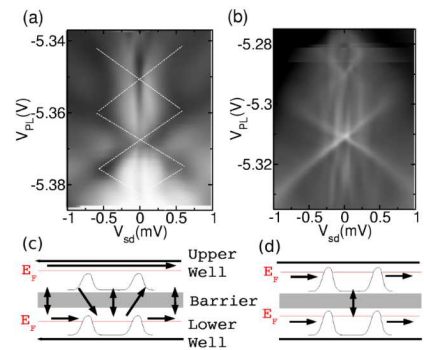}
\caption{\label{fig2:epsart} (Color online) Differential conductance, dI/dV$_{sd}$ measured as a function of V$_{PL}$ and V$_{sd}$ in the region of (a) Coulomb oscillations and (b) Fano dips. V$_{TOP}$=-0.05V. The greyscale corresponds to white = 0$\mu$S(50$\mu$S) to black = 65$\mu$S(70$\mu$S) for (a) and (b) respectively. Note that the range of V$_{PL}$ does not match Fig.1 (c) or Fig. 3(a) due to switching events occuring between the measurements. Dashed white lines in (a) are a guide to the eye. (c) A schematic diagram showing the Fermi energy in each well in relation to the barriers created by V$_{L1}$ and V$_{L2}$. Arrows indicate possible electron paths which could give rise to the Fano dips. (d) The situation when a dot is formed in each well.}
\end{figure}

\begin{figure}[t]
\includegraphics{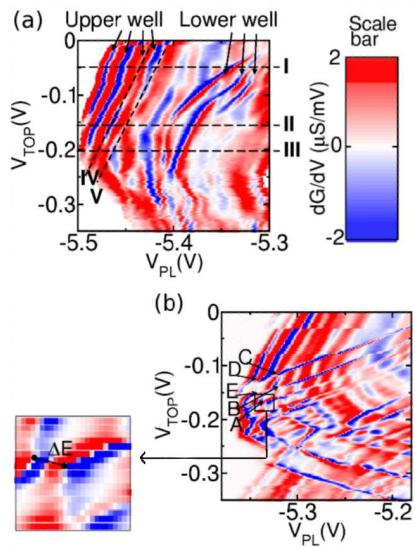}
\caption{\label{fig3:epsart} dG/dV$_{PL}$ plotted as a function of V$_{PL}$ and V$_{TOP}$ for (a) B=0T and (b) B=5T in the plane of the 2DEG's. Lines I-V are explained in the text. The scale bar corresponds to (a) and (b).}
\end{figure}

The splitting of the triple points at five different gate voltages has been measured as a function of magnetic field (Fig. 4(a)). The axis is calibrated in terms of energy from the measurements made in source-drain bias (Fig. 2). The effect of the magnetic field is to confine the electronic wavefunction in each well and reduce the overlap of the wavefunction between the wells, thereby reducing t. This has been predicted theoretically\cite{Burkard} and observed in columnar mesas where the electron number is very small\cite{Ancilotto}. An exact analytical form for t(B) is not available when the electron number is high, as in our case, but we note that the general behaviour is for the splitting to decrease and saturate at B$\approx$5T. This corresponds to a magnetic length, l$_{B}$=($\hbar$/eB)$^{1/2}$=11nm. The centre to centre distance of the quantum wells is 18.5nm which is approximately 2l$_{B}$, so the splitting at high fields is due to the capacitive coupling between the dots when there is no overlap of the wavefunction and t=0. The saturation at low fields (B$<$3T) occurs when the coupling is strong and the total splitting becomes comparable to the dot charging energy, U or t$\approx$$\delta$E\cite{DeltaE} and transitions of the ground state from bonding to antibonding and vice-versa can occur\cite{Ancilotto}. The size of the splitting will depend on the individual states involved. Typically t$>$100$\mu$eV. 
\begin{figure}[t]
\includegraphics{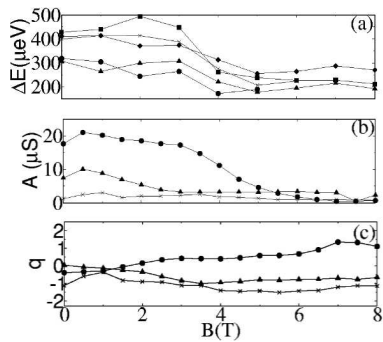}
\caption{\label{fig4:epsart}(a) The magnetic field dependence of the splittings measured at points A-E in Fig. 3(b) corresponding to circles, squares, triangles, diamonds and crosses respectively. (b)-(c) The dependence of the fitting parameters in Eq. 1 for the dips labelled 1 (circles), 2 (triangles) and 3 (crosses) in Fig. 1(c).}
\end{figure}

The magnetic field dependence of the Fano parameters (Fig. 4(b)-(c)) are consistent with the argument that the interfering paths are in separate, coupled quantum wells. The general decrease of A and the increase of $\mid$q$\mid$$\propto$V$_{E}$$^{-1}$ reflect the decrease in the overlap between the continuum and resonant states. A falls towards zero at approximately the same magnetic field as the saturation of $\Delta$E. In our analysis we have been able to fit Eq.1 by treating q as a real number. This is in contrast to the analysis for a dot embedded in one arm of an Aharonov-Bohm ring\cite{Kobayashi} where a complex q was required to account for the breaking of time reversal symmetry as the magnetic flux threaded the ring. The success of our treatment indicates that the loop formed by the current paths in our device is small in comparison \cite{signchange}. Alternatively, the geometry may resemble a stub resonator \cite{Kobayashi2}. The possible electron paths giving rise to the Fano dips are shown in Fig. 2(c).

\begin{figure}[t]
\includegraphics{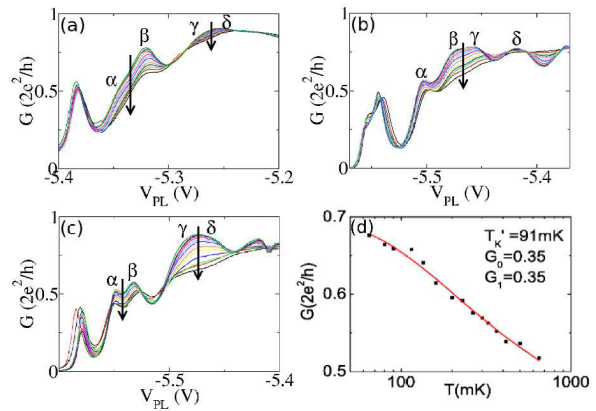}
\caption{\label{fig5:epsart} (Color online) (a)-(c) The temperature dependence of the conductance for values of V$_{TOP}$ indicated by lines I,II and III respectively in Fig. 3(a). Arrows show the trend for increasing temperature. (d) An example of the logarithmic temperature dependence of the conductance between peaks $\alpha$ and $\beta$ in (a). The parameters are explained in the text.}
\end{figure}

We also observe evidence for the Kondo effect in our device. Figure 5 (a)-(c) shows the temperature dependence of the conductance for three different gate voltages applied to V$_{TOP}$. In Fig. 5(a) the conductance in the valleys between peaks $\alpha$ and $\beta$ is enhanced at low temperatures. This is also the case between peaks $\gamma$ and $\delta$, but not between peaks $\beta$ and $\gamma$. The regions of enhanced conductance exhibit a logarithmic temperature dependence (for example see Fig. 5(d)) and a peak in dI/dV at zero source-drain bias as observed in Fig. 2(a)\cite{ZBA}. This behaviour is characteristic of the Kondo effect which arises due to the presence of an unpaired spin in a degenerate state coupling to the electrons in the leads. This indicates that the dot in the upper well is alternating between unpaired(U), then paired(P), then unpaired spin states as the voltage is swept between peaks $\alpha$ and $\delta$. In Fig. 5(b) the order of this alternating behaviour is reversed with the region of enhanced conductance now being observed between peaks $\beta$ and $\gamma$ indicating that this valley now represents the dot in the upper well containing an unpaired spin. In Fig. 5(c) the ordering is restored to that observed in Fig. 5(a). The fact that the order of the regions of enhanced conductance changes between Figs. 5(a)-(c) is consistent with an electron being transferred from the dot in the upper well to the dot in the lower well as V$_{TOP}$ is made more negative. Following line IV in Fig. 3(a) this could represent the states between peaks $\alpha$ and $\beta$ undergoing the transitions (U,U) $\rightarrow$ (P,P) $\rightarrow$ (U,U) as V$_{TOP}$ is made more negative. Here the notation in brackets represents the spin states of the (upper,lower) dots.  Similarly, line V could represent the states between peaks $\beta$ and $\gamma$ undergoing the transitions (P,U) $\rightarrow$ (U,P) $\rightarrow$ (P,U). We find that the regions of enhanced conductance are fitted well by G$=$G$_0$$+$G$_1$(T$^\prime$$_K$$^2$/(T$^\prime$$_K$$^2$+T$^2$))$^s$ (for example, the red curve in Fig. 5(d)). This is the empirical formula used to describe the conductance in the Kondo regime, where s$=$0.2 for a spin 1/2 system and T$^\prime$$_K$$=$T$_K$(2$^{1/s}$-1)$^{-1/2}$, where T$_K$ is the Kondo temperature\cite{Goldhaber2}. G$_0$ is necessary in our case to account for a background due to co-tunneling. Generally we find s$=$0.2, but deviates strongly from this value between peaks $\gamma$ and $\delta$. This deviation, and the finite value of G$_0$ is likely due to strong coupling to the leads, which can lead to deviations from the Anderson impurity model\cite{Glazman}.  

In conclusion, the lateral transport through strongly, vertically coupled quantum dots reveals both the Fano effect and the Kondo effect. The observation of both of these effects in one device offers the possibility of studying quantum interference and many body effects in the same system and may find practical applications in quantum computation.

\begin{acknowledgments}
We gratefully acknowledge funding by the EPSRC (UK), and discussions facilitated by the EU COLLECT network.
\end{acknowledgments}



\begin{thebibliography}{}
\bibitem{electrontemp} We estimate the electron temperature to be approximately 80mK from the width of Coulomb oscillations as a function of temperature.
\bibitem{Isolated} In this region of gate voltage there could still be charge in the lower well, but conductance through that well is blocked by large entrance and exit barriers.
\bibitem{Explain shift} The data in Fig. 3 has been shifted along the x-axis by 4mV for every 1mV increase in V$_{TOP}$. This is to allow the features to be seen clearly, otherwise they would be spread over a 1.4V range and the different features would not be easily distinguishable to the eye. Therefore, it is not clear from Fig. 3 that the Coulomb oscillations actually have a larger capacitance to V$_{TOP}$ than do the Fano features, as would be expected considering the closer proximity of the upper well to gate TOP and the screening of the lower well.
\bibitem{DeltaE} From the charging energy in Fig. 2(a), we estimate that the lower dot contains approximately 100 electrons by modelling it as a disc with diameter D and capacitance C=4$\pi$$\epsilon$$_{r}$$\epsilon$$_{0}$D. This corresponds to a single particle level spacing, $\delta$E$\approx$90$\mu$eV.
\bibitem{signchange} q appears to change sign at approximately 1T. This may indicate that there is some breaking of time reversal symmetry, but that the loop formed is very small in comparison to reference \cite{Kobayashi}.
\bibitem{ZBA}  Zero bias peaks are also observed in some valleys which do not exhibit the strong temperature dependence. This may be caused by an unpaired spin on the lower dot being weakly coupled to the leads.
\end{thebibliography}

\end{document}